\documentclass[aps,showpacs,twocolumn]{revtex4}
\usepackage{epsfig}
\usepackage{graphicx,color,dcolumn}
\usepackage{epstopdf}
\usepackage{amsmath,amssymb}
\usepackage{multirow}
\usepackage{diagbox}
\usepackage{changes}
\usepackage{booktabs}
\usepackage{threeparttable}

\begin{document}

\title{ $Y(4626)$ in a chiral constituent quark model}

\author{Yue Tan}\email{181001003@stu.njnu.edu.cn}
\author{Jialun Ping}\email{jlping@njnu.edu.cn(Corresponding author)}

\affiliation{Department of Physics and Jiangsu Key Laboratory for Numerical
Simulation of Large Scale Complex Systems, Nanjing Normal University, Nanjing 210023, P. R. China}

\begin{abstract}
Recently, Belle Collaboration reported a new exotic state $Y(4620)$ with mass at 4625.9 MeV in the positronium annihilation process.
Inspired by experiment, we study the tetraquark system $c\bar{s}s\bar{c}$ with quantum numbers $J^{P}=1^{-}$ in the framework
of chiral constituent quark model with the help of Gaussian expansion method. Two structures, diquark-antidiquark and
meson-meson, with all possible color and spin configurations are considered. The result shows that no bound state can be
formed. To investigate the possible resonance states, the real scaling method is employed. Several resonance states with
energies 4354, 4408, 4469, 4497 and 4531 MeV, are proposed. Taking into account the errors in calculating the $q\bar{q}$ mesons,
the system errors in the calculation of four-quark system are around 60$\sim 100$ MeV. The resonance with energy 4531 MeV
is possible the candidate of the newly found state $Y(4620)$.
\end{abstract}

\pacs{~}

\maketitle

\section{Introduction} \label{introduction}

Since the state $X(3872)$ first observed by the Belle collaboration~\cite{01Choi:2003ue}, a lot of new hadrons have been reported
subsequently the other
collaborations~\cite{02Abe:2004zs,03Aubert:2005rm,04Choi:2007wga,05Yuan:2007sj,06Mizuk:2008me,07Aaltonen:2009tz,08Aaij:2019evc}.
Most of them cannot be fitted well in the conventional picture of meson and baryon, which are called exotic states.
In fact. the exotic states can be divided into two kinds: the first kind is the states with exotic quantum numbers, and the second one
is the states with normal quantum numbers but their properties cannot be described by the conventional quark models. In the picture
of quark model, the meson is made up of quark-antiquark and baryon is made up of three quarks. With the accumulation of experimental
data on exotic states, people believe that these exotic states can provide much essential information on low energy QCD and help us
to establish the effective method to describe all hadrons.

Very recently, the Belle Collaboration observed a new structure, named $Y(4626)$, in the $D_sD_{s1}(2536)$ invariant mass spectrum
with 5.9$\sigma$ significance \cite{00Jia:2019gfe}. The mass and width measured is $M = 4265.9^{+6.2}_{-6.0}\pm0.4$ MeV and
$\Gamma = 49.8^{+13.9}_{-11.5}\pm 4.0 $ MeV, respectively. Its decay mode $D_sD_{s1}(2536)$ indicates that $Y(4626)$ is consist of
charm and strange quarks. The Belle Collaboration also suggested that the quantum numbers of the state is $J^{P}=1^{-}$.
Because the mass is close to the threshold of $D_s+D_{s1}(2536)$, the possible assignment is $c\bar{s}s\bar{c}$ four-quark state.

There were lots of researches about $c\bar{s}s\bar{c}$ before \cite{09Chen:2016oma,10Deng:2017xlb,11Yang:2019dxd,12Lu:2016cwr,13Wu:2016gas,14Stancu:2009ka,15Ortega:2016hde}.
For example, Chen {\em et al.} analyzed the tetraquark states $c\bar{s}s\bar{c}$ with $J^{P}=1^{+}$ (in $S$-wave) and
$J^{P}=0^{+}$ (in $D$-wave) within within the framework of QCD sum rules~\cite{09Chen:2016oma}. Deng {\em et al.} used
color flux-tube model to investigate systematically the hidden charmed states observed in recent years including
$J^{P}=0^{+}$ and $J^{P}=1^{+}$ $c\bar{s}s\bar{c}$ system \cite{10Deng:2017xlb}. Recently, in the framework of the chiral
quark model Yang {\em et al.} investigated the four-quark system $c\bar{s}s\bar{c}$ with quantum numbers $1^{+}$ and $0^{+}$,
and described the $X(4274)$ and $X(4350)$ states in diquark-antidiquark picture~\cite{11Yang:2019dxd}. So far, most of the
previous work considered only the four-quark states with negative parity, the four-quark states with $J^{P}=1^{-}$ is
nevertheless untouched.

In this paper, a constituent quark model is employed to systematically investigate the $c\bar{s}s\bar{c}$ states with $J^{P}=1^{-}$
with the help of gaussian expansion method. In the calculation, all of possible color and spin configurations are considered.
In addition, two structures, meson-meson and diquark-antidiquark and their mixing, are also taken into account.
In fact, one structure is complete for the calculation, if all excitation of the structure are taken into account. Clearly
it is too difficult to use this approach. An economic way is to combine different structures, which are kept in the low-lying
state to do the calculation. In this approach, the problem of over-complete will shown up. To solve the problem, the
eigenfunction method is employed. First the overlap matrix is diagonized, the eigenvectors with eigenvalue 0 are
abandoned, then re-constructed the hamiltonian matrix using the remained eigenvectors of overlap matrix. At last, the new hamiltonian
matrix is diagonized to obtained the eigen-energies of the system. In order to keep the matrix manageable, the important structures,
meson-meson and diquark-antidiquark structures which are favorable physically, are considered. Other structures, e.g., K-type structure
are not favorable, so they are not taken into account temporarily.

The paper is organized as follows. In Sect. II, the chiral quark model and the wave functions of the $c\bar{s}s\bar{c}$ with
quantum numbers $J^{P}=1^{-}$ are presented. The numerical results are given in Sec. III.
The last section is devoted to the summary of the present work.

\section{Chiral quark model and wave function of $c\bar{s}s\bar{c}$ system} \label{wavefunction and chiral quark model}
\subsection{Chiral quark model}
The chiral quark model has been applied successfully in describing the hadron spectra and hadron-hadron interactions.
The details of the model can be found in Ref. \cite{16Vijande:2004he,17Yang:2008zzi,18Yang:2009zzp,19Chen:2016npt,20Chen:2017mug,21Tan:2019qwe}.
Here only the Hamiltonian of the chiral quark model for four-quark system is shown,
\begin{eqnarray}
H &=& \sum_{i=1}^4
m_i+\frac{p_{12}^2}{2\mu_{12}}+\frac{p_{34}^2}{2\mu_{34}}
  +\frac{p_{1234}^2}{2\mu_{1234}}  \nonumber \\
  &+ & \sum_{i<j=1}^4 \left( V_{ij}^{C}+V_{ij}^{G}+\sum_{\chi=\pi,K,\eta, \sigma} V_{ij}^{\chi} \right),
\end{eqnarray}
where $m_i$ is the constituent masse of $i$-th quark (antiquark), and $\mu$ is the reduced masse of two interacting quarks
or quark-clusters.
\begin{eqnarray}
\mu_{ij}&=&\frac{m_{{i}}  m_{{j}}}{m_{{i}} + m_{{j}}}, ~~ij=12,34   \nonumber\\
\mu_{1234}&=&\frac{(m_1+m_2)(m_3+m_4)}{m_1+m_2+m_3+m_4},\\
p_{ij}&=&\frac{m_jp_i-m_ip_j}{m_i+m_j}, \nonumber \\
p_{1234}&=&\frac{(m_3+m_4)p_{12}-(m_1+m_2)p_{34}}{m_1+m_2+m_3+m_4}.\nonumber
\end{eqnarray}
$V^C$ is the confining potential, mimics the ``confinement" property of QCD,
\begin{equation}
V_{ij}^{C} = ( -a_{c} r_{ij}^{2}-\Delta) \boldsymbol{\lambda}_i^c \cdot \boldsymbol{\lambda}_j^c
\end{equation}
The second potential $V^{G}$ is one-gluon exchange interaction reflecting the ``asymptotic freedom" property of QCD.
\begin{eqnarray}
 V_{ij}^{G}&=& \frac{\alpha_s}{4} \boldsymbol{\lambda}_i^c \cdot \boldsymbol{\lambda}_{j}^c
\left[\frac{1}{r_{ij}}-\frac{2\pi}{3m_im_j}\boldsymbol{\sigma}_i\cdot
\boldsymbol{\sigma}_j
  \delta(\boldsymbol{r}_{ij})\right]   \\
  & &  \delta{(\boldsymbol{r}_{ij})}=\frac{e^{-r_{ij}/r_0(\mu_{ij})}}{4\pi r_{ij}r_0^2(\mu_{ij})}, \nonumber
\end{eqnarray}
$\boldsymbol{\sigma}$ are the $SU(2)$ Pauli matrices; $\boldsymbol{\lambda}_{c}$ are $SU(3)$ color Gell-Mann matrices,
$r_{0}(\mu_{ij})=\frac{r_0}{\mu_{ij}}$ and $\alpha_{s}$ is an effective scale-dependent running coupling,
\begin{equation}
 \alpha_s(\mu_{ij})=\frac{\alpha_0}{\ln\left[(\mu_{ij}^2+\mu_0^2)/\Lambda_0^2\right]}.
\end{equation}
The third potential $V_{\chi}$ is Goldstone boson exchange, coming from ``chiral symmetry spontaneous breaking" of QCD
in the low-energy region,
\begin{eqnarray}
V_{ij}^{\pi}&=& \frac{g_{ch}^2}{4\pi}\frac{m_{\pi}^2}{12m_im_j}
  \frac{\Lambda_{\pi}^2}{\Lambda_{\pi}^2-m_{\pi}^2}m_\pi v_{ij}^{\pi}
  \sum_{a=1}^3 \lambda_i^a \lambda_j^a,  \nonumber \\
V_{ij}^{K}&=& \frac{g_{ch}^2}{4\pi}\frac{m_{K}^2}{12m_im_j}
  \frac{\Lambda_K^2}{\Lambda_K^2-m_{K}^2}m_K v_{ij}^{K}
  \sum_{a=4}^7 \lambda_i^a \lambda_j^a,  \nonumber \\
V_{ij}^{\eta} & = &
\frac{g_{ch}^2}{4\pi}\frac{m_{\eta}^2}{12m_im_j}
\frac{\Lambda_{\eta}^2}{\Lambda_{\eta}^2-m_{\eta}^2}m_{\eta}
v_{ij}^{\eta}  \\
 && \left[\lambda_i^8 \lambda_j^8 \cos\theta_P
 - \lambda_i^0 \lambda_j^0 \sin \theta_P \right],  \nonumber \\
V_{ij}^{\sigma}&=& -\frac{g_{ch}^2}{4\pi}
\frac{\Lambda_{\sigma}^2}{\Lambda_{\sigma}^2-m_{\sigma}^2}m_\sigma
\left[
 Y(m_\sigma r_{ij})-\frac{\Lambda_{\sigma}}{m_\sigma}Y(\Lambda_{\sigma} r_{ij})\right], \nonumber \\
 v_{ij}^{\chi} & = & \left[ Y(m_\chi r_{ij})-
\frac{\Lambda_{\chi}^3}{m_{\chi}^3}Y(\Lambda_{\chi} r_{ij})
\right]
\boldsymbol{\sigma}_i \cdot\boldsymbol{\sigma}_j,~~ \chi=\pi,K,\eta, \nonumber \\
& & Y(x)  =   e^{-x}/x. \nonumber
\end{eqnarray}
$\boldsymbol{\lambda}$ are $SU(3)$ flavor Gell-Mann matrices, $m_{\chi}$ are the masses of Goldstone bosons,
$\Lambda_{\chi}$ are the cut-offs, $g^2_{ch}/4\pi$ is the Goldstone-quark coupling constant.

All the parameters are determined by fitting the meson spectrum, from light to heavy, taking into account only a
quark-antiquark component. They are shown in Table~\ref{modelparameters}. The calculated masses of the mesons involved
in the present work are shown in Table~\ref{mesonmass}. Because the spin-orbit interaction is not considered here,
we obtained a degenerate eigen-energy for the three $P$-wave states, $^3P_J,J=0,1,2$.

\begin{table}[t]
\begin{center}
\caption{Quark model parameters ($m_{\pi}=0.7$ fm, $m_{\sigma}=3.42$ fm, $m_{\eta}=2.77$ fm, $m_{K}=2.51$ fm).\label{modelparameters}}
\begin{tabular}{cccc}
\hline\hline\noalign{\smallskip}
Quark masses   &$m_u=m_d$(MeV)     &313  \\
               &$m_{s}$(MeV)         &536  \\
               &$m_{c}$(MeV)         &1728 \\
               &$m_{b}$(MeV)         &5112 \\
\hline
Goldstone bosons   
                   &$\Lambda_{\pi}=\Lambda_{\sigma}(fm^{-1})$     &4.2  \\
                   &$\Lambda_{\eta}=\Lambda_{K}(fm^{-1})$     &5.2  \\
                   &$g_{ch}^2/(4\pi)$                &0.54  \\
                   &$\theta_p(^\circ)$                &-15 \\
\hline
Confinement             &$a_{c}$(MeV)     &101 \\
                   &$\Delta$(MeV)       &-78.3 \\
                   &$\mu_{c}$(MeV)       &0.7 \\
\hline
OGE                 & $\alpha_{0}$        &3.67 \\
                   &$\Lambda_{0}(fm^{-1})$ &0.033 \\
                  &$\mu_0$(MeV)    &36.976 \\
                   &$\hat{r}_0$(MeV)    &28.17 \\
\hline\hline
\end{tabular}
\end{center}
\end{table}

\begin{table}[]
\caption{ \label{mesonmass} Meson spectrum (unit: MeV).}
\begin{tabular}{cccccc}
\hline\noalign{\smallskip}
                            &     & $^{1}S_{0}$ & $^{3}S_{1}$ & $^{1}P_{1}$ & $^{3}P_{J}$ \\ \hline
\multirow{2}{*}{$c\bar{c}$} & QM  & ~~2986.3~~      & ~~3096.4~~      & ~~3416.3~~      & ~~3417.2~~ \\
                            & ~~PDG~~ & 2979.6      & 3096.9      & 3526.2      & 3510.6 \\ \hline
\multirow{2}{*}{$c\bar{s}$} & QM  & 1953.3      & 2080.6      & 2479.3      & 2482.9 \\
                            & PDG & 1981.0      & 2112.0      & 2460.0      & 2536.0 \\ \hline
\multirow{2}{*}{$s\bar{s}$} & QM  & 824.0       & 1015.8      & 1469.1      & 1481.3 \\
                            & PDG & 957.8       & 1019.4      & 1386.0      & 1426.3 \\
\hline
\end{tabular}
\end{table}

\subsection{The wave function of $c\bar{s}s\bar{c}$ system}
There are two physically important structures, meson-meson and diquark-antidiquark, are considered in the present calculation.
The wave functions of every structure all consists of four parts: orbital, spin, flavor
and color. The wave function of each part is constructed in two steps, first write down the two-body wave functions, then coupling
two sub-clusters wave functions to form the four-body one. Because there is no identical particles in the system, the total wave
function of the system is the direct product of orbital ($|R_{i}\rangle$), spin ($|S_{j}\rangle$), color ($|C_{k}\rangle$) and
flavor ($|F_{n}\rangle$) wave functions with necessary coupling,
\begin{equation}\label{bohanshu}
|ijkn\rangle=[|R_{i}\rangle\otimes|S_{j}\rangle]\otimes|C_{k}\rangle\otimes |F_{n}\rangle
\end{equation}

\subsubsection{orbital wave function}
The orbital wave function of the four-quark system consists of two sub-cluster orbital wave function and the relative
motion wave function between two subclusters (1,3 denote quarks and 2,4 denote antiquarks),
\begin{eqnarray}\label{spatialwavefunctions}
|R_{1}\rangle&=&\left[[\Psi_{l_1=1}({\bf r}_{12})\Psi_{l_2=0}({\bf
r}_{34})]_{l_{12}}\Psi_{L_r}({\bf r}_{1234}) \right]_{L}^{M_{L}}, \nonumber\\
|R_{2}\rangle&=&\left[[\Psi_{l_1=0}({\bf r}_{12})\Psi_{l_2=1}({\bf
r}_{34})]_{l_{12}}\Psi_{L_r}({\bf r}_{1234}) \right]_{L}^{M_{L}}, \nonumber\\
|R_{3}\rangle&=&\left[[\Psi_{l_1=1}({\bf r}_{13})\Psi_{l_2=0}({\bf
r}_{24})]_{l_{12}}\Psi_{L_r}({\bf r}_{1324}) \right]_{L}^{M_{L}}, \\
|R_{4}\rangle&=&\left[[\Psi_{l_1=0}({\bf r}_{13})\Psi_{l_2=1}({\bf
r}_{24})]_{l_{12}}\Psi_{L_r}({\bf r}_{1324}) \right]_{L}^{M_{L}}, \nonumber
\end{eqnarray}
where the bracket "[~]" indicates orbital angular momentum coupling, and $L$ is the total orbital angular momentum which
comes from the coupling of $L_r$, orbital angular momentum of relative motion, and $l_{12}$, which coupled by $l_1$ and $l_2$,
sub-cluster orbital angular momenta. $|R_{1}\rangle, |R_{2}\rangle$ donate the orbital wave functions of meson-meson structure,
and $|R_{3}\rangle, |R_{4}\rangle$ donate the wave functions of diquark-antidiquark structure.
In GEM, the radial part of the orbital wave function is expanded by a set of Gaussians:
\begin{subequations}
\label{radialpart}
\begin{align}
\Psi(\mathbf{r}) & = \sum_{n=1}^{n_{\rm max}} c_{n}\psi^G_{nlm}(\mathbf{r}),\\
\psi^G_{nlm}(\mathbf{r}) & = N_{nl}r^{l}
e^{-\nu_{n}r^2}Y_{lm}(\hat{\mathbf{r}}),
\end{align}
\end{subequations}
where $N_{nl}$ are normalization constants,
\begin{align}
N_{nl}=\left[\frac{2^{l+2}(2\nu_{n})^{l+\frac{3}{2}}}{\sqrt{\pi}(2l+1)}
\right]^\frac{1}{2}.
\end{align}
$c_n$ are the variational parameters, which are determined dynamically. The Gaussian size parameters are chosen according
to the following geometric progression
\begin{equation}\label{gaussiansize}
\nu_{n}=\frac{1}{r^2_n}, \quad r_n=r_1a^{n-1}, \quad
a=\left(\frac{r_{n_{\rm max}}}{r_1}\right)^{\frac{1}{n_{\rm max}-1}}.
\end{equation}
This procedure enables optimization of the using of Gaussians, as small as possible Gaussians are used.

\subsubsection{spin wave function}
Because of no difference between spin of quark and antiquark, the meson-meson structure has the same spin wave function as
the diquark-antidiquark structure. The spin wave functions of the sub-cluster are shown below.
\begin{align*}
&\chi_{11}^{\sigma}=\alpha\alpha,~~
\chi_{10}^{\sigma}=\frac{1}{\sqrt{2}}(\alpha\beta+\beta\alpha),~~
\chi_{1-1}^{\sigma}=\beta\beta,\nonumber \\
&\chi_{00}^{\sigma}=\frac{1}{\sqrt{2}}(\alpha\beta-\beta\alpha),
\end{align*}
Coupling the spin wave functions of two sub-clusters by Clebsch-Gordan coefficients, total spin wave function can be written below,
\begin{align*}
|S_{1}\rangle=\chi_{0}^{\sigma1}&=\chi_{00}^{\sigma}\chi_{00}^{\sigma},\\
|S_{2}\rangle=\chi_{0}^{\sigma2}&=\sqrt{\frac{1}{3}}(\chi_{11}^{\sigma}
  \chi_{1-1}^{\sigma}-\chi_{10}^{\sigma}\chi_{10}^{\sigma}+\chi_{1-1}^{\sigma}\chi_{11}^{\sigma}),\\
|S_{3}\rangle=\chi_{1}^{\sigma1}&=\chi_{00}^{\sigma}\chi_{11}^{\sigma},\\
|S_{4}\rangle=\chi_{1}^{\sigma2}&=\chi_{11}^{\sigma}\chi_{00}^{\sigma},\\
|S_{5}\rangle=\chi_{1}^{\sigma3}&=\frac{1}{\sqrt{2}}(\chi_{11}^{\sigma}\chi_{10}^{\sigma}-\chi_{10}^{\sigma}\chi_{11}^{\sigma}),\\
|S_{6}\rangle=\chi_{2}^{\sigma1}&=\chi_{11}^{\sigma}\chi_{11}^{\sigma}.\\
\end{align*}
the total spin wave function is denoted by $\chi_{S}^{\sigma i}$, $i$ is the index of the functions, the $S$ is the total spin
of the system.

\subsubsection{flavor wave function}
We have two flavor wave functions of the system,
\begin{align*}
|F_{1}\rangle&=(c\bar{s})(s\bar{c}),\\
|F_{2}\rangle&=(s\bar{s})(c\bar{c}),\\
|F_{3}\rangle&=(cs)(\bar{c}\bar{s}).
\end{align*}
$|F_{1}\rangle, |F_{2}\rangle$ is for meson-meson structure, and $|F_{3}\rangle$ is for diquark-antidiquark structure.

\subsubsection{color wave function}
The colorless tetraquark system has four color wave functions, two for meson-meson structure, $1\otimes1$ ($C_1$), $8\otimes8$ ($C_2$),
and two for diquark-antidiquark structure, $\bar{3}\otimes 3$ ($C_3$) and $6\otimes \bar{6}$ ($C_4$).
\begin{eqnarray}
|C_{1}\rangle &= & \sqrt{\frac{1}{9}}(\bar{r}r\bar{r}r+\bar{r}r\bar{g}g+\bar{r}r\bar{b}b+\bar{g}g\bar{r}r+\bar{g}g\bar{g}g \nonumber\\
 & & +\bar{g}g\bar{b}b+\bar{b}b\bar{r}r+\bar{b}b\bar{g}g+\bar{b}b\bar{b}b), \nonumber \\
|C_{2}\rangle & = & \sqrt{\frac{1}{72}}(3\bar{b}r\bar{r}b+3\bar{g}r\bar{r}g+3\bar{b}g\bar{g}b+3\bar{g}b\bar{b}g+3\bar{r}g\bar{g}r \nonumber \\
& & +3\bar{r}b\bar{b}r+2\bar{r}r\bar{r}r+2\bar{g}g\bar{g}g+2\bar{b}b\bar{b}b-\bar{r}r\bar{g}g \nonumber \\
& & -\bar{g}g\bar{r}r-\bar{b}b\bar{g}g-\bar{b}b\bar{r}r-\bar{g}g\bar{b}b-\bar{r}r\bar{b}b). \\
|C_{3}\rangle &= &
 \sqrt{\frac{1}{12}}(rg\bar{r}\bar{g}-rg\bar{g}\bar{r}+gr\bar{g}\bar{r}-gr\bar{r}\bar{g}+rb\bar{r}\bar{b} \nonumber \\
 & & -rb\bar{b}\bar{r}+br\bar{b}\bar{r}-br\bar{r}\bar{b}+gb\bar{g}\bar{b}-gb\bar{b}\bar{g} \nonumber \\
 & & +bg\bar{b}\bar{g}-bg\bar{g}\bar{b}). \nonumber \\
|C_{4}\rangle &= & \sqrt{\frac{1}{24}}(2rr\bar{r}\bar{r}+2gg\bar{g}\bar{g}+2bb\bar{b}\bar{b}
    +rg\bar{r}\bar{g}+rg\bar{g}\bar{r} \nonumber \\
& & +gr\bar{g}\bar{r}+gr\bar{r}\bar{g}+rb\bar{r}\bar{b}+rb\bar{b}\bar{r}+br\bar{b}\bar{r} \nonumber \\
& & +br\bar{r}\bar{b}+gb\bar{g}\bar{b}+gb\bar{b}\bar{g}+bg\bar{b}\bar{g}+bg\bar{g}\bar{b}).
\end{eqnarray}

\begin{table}[tp]
\centering
\caption{Index of physical channels for $c\bar{s}s\bar{c}$ system.}\label{channel}
\begin{tabular}{cccc}
\hline \hline
$|ijkn\rangle$ & $S=0$ & $|ijkn\rangle$ & $S=1$ \\ \hline
$|1111\rangle$&$D_s\bar{D}_{s1}(2460)$               &$|1311\rangle$&$D_s\bar{D}_{sj}$                     \\
$|1121\rangle$&~~$[D_s]_8[\bar{D}_{s1}(2460)]_8$~~   &$|1321\rangle$&$[D_s]_8[\bar{D}_{sj}]_8$             \\
$|1211\rangle$&$D_s^{*}\bar{D}_{sj}$                 &$|1411\rangle$&$D_s^*\bar{D}_{s1}(2460)$             \\
$|1221\rangle$&$[D_s^{*}]_8[\bar{D}_{sj}]_8$         &$|1421\rangle$&~~$[D_s^*]_8[\bar{D}_{s1}(2460)]_8$~~     \\
$|2111\rangle$&$D_{s1}(2460)\bar{D}_{s}$             &$|1511\rangle$&$D_s^*\bar{D}_{sj}$                   \\
$|2121\rangle$&$[D_{s1}(2460)]_8[\bar{D}_{s}]_8$     &$|1521\rangle$&$[D_s^*]_8[\bar{D}_{sj}]_8$           \\
$|2211\rangle$&$D_{sj}\bar{D}_{s}^*$                 &$|2311\rangle$&$D_{s1}(2460)\bar{D}_s^*$             \\
$|2221\rangle$&$[D_{sj}]_8[\bar{D}_{s}^*]_8$         &$|2321\rangle$&$[D_{s1}(2460)]_8[\bar{D}_s^*]_8$     \\
$|1112\rangle$&${\eta}^{\prime}h_c$                  &$|2411\rangle$&$D_{sj}\bar{D}_s$                     \\
$|1122\rangle$&$[{\eta}^{\prime}]_8[h_c]_8$          &$|2421\rangle$&$[D_{sj}]_8[\bar{D}_s]_8$             \\
$|1212\rangle$&${\phi}^{\prime}\chi_{cj}$            &$|2511\rangle$&$D_{sj}\bar{D}_s^*$                   \\
$|1222\rangle$&$[{\phi}^{\prime}]_8[\chi_{cj}]_8$    &$|2521\rangle$&$[D_{sj}]_8[\bar{D}_s^*]_8$           \\
$|2112\rangle$&${h\eta_c}$                           &$|1312\rangle$&${\eta}^{\prime}\chi_{cj}$                 \\
$|2122\rangle$&${[h]_8[\eta_c]_8}$                   &$|1322\rangle$&$[{\eta}^{\prime}]_8[\chi_{cj}]_8$         \\
$|2212\rangle$&$f_jJ/\psi$                           &$|1412\rangle$&${\phi}h_{c}$                         \\
$|2222\rangle$&$[f_j]_8[J/\psi]_8$                   &$|1422\rangle$&$[{\phi}]_8[h_{c}]_8$                 \\
$|3133\rangle$&$[sc]_3^0[\bar{s}\bar{c}]_{\bar{3}}^0$&$|1512\rangle$&${\phi}\chi_{cj}$                     \\
$|3143\rangle$&$[sc]_6^0[\bar{s}\bar{c}]_{\bar{6}}^0$&$|1522\rangle$&$[{\phi}]_8[\chi_{cj}]_8$             \\
$|3233\rangle$&$[sc]_3^1[\bar{s}\bar{c}]_{\bar{3}}^1$&$|2312\rangle$&$hJ/\psi$                             \\
$|3243\rangle$&$[sc]_6^1[\bar{s}\bar{c}]_{\bar{6}}^1$&$|2322\rangle$&$[h]_8[J/\psi]_8$                     \\
$|4133\rangle$&$[sc]_3^0[\bar{s}\bar{c}]_{\bar{3}}^0$&$|2412\rangle$&$f_j\eta_c$                           \\
$|4143\rangle$&$[sc]_6^0[\bar{s}\bar{c}]_{\bar{6}}^0$&$|2422\rangle$&$[f_j]_8[\eta_c]_8$                   \\
$|4233\rangle$&$[sc]_3^1[\bar{s}\bar{c}]_{\bar{3}}^1$&$|2512\rangle$&$f_jJ/\psi$                           \\
$|4243\rangle$&$[sc]_6^1[\bar{s}\bar{c}]_{\bar{6}}^1$&$|2522\rangle$&$[f_j]_8[J/\psi]_8$                   \\ \cline{1-2}
$|ijkn\rangle$ & $S=2$                               &$|3333\rangle$&$[sc]_3^0[\bar{s}\bar{c}]_{\bar{3}}^1$\\ \cline{1-2}
$|1611\rangle$&$D_s^*\bar{D}_{sj}$                   &$|3343\rangle$&$[sc]_6^0[\bar{s}\bar{c}]_{\bar{6}}^1$\\
$|1621\rangle$&$[D_s^*]_8[\bar{D}_{sj}]_8$           &$|3433\rangle$&$[sc]_3^1[\bar{s}\bar{c}]_{\bar{3}}^0$\\
$|2611\rangle$&$D_{sj}\bar{D}_s^*$                   &$|3443\rangle$&$[sc]_6^1[\bar{s}\bar{c}]_{\bar{6}}^0$\\
$|2621\rangle$&$[D_{sj}]_8[\bar{D}_s^*]_8$           &$|3533\rangle$&$[sc]_3^1[\bar{s}\bar{c}]_{\bar{3}}^1$\\
$|1612\rangle$&${\phi}\chi_{cj}$                     &$|3543\rangle$&$[sc]_6^1[\bar{s}\bar{c}]_{\bar{6}}^1$\\
$|1622\rangle$&$[{\phi}]_8[\chi_{cj}]_8$             &$|4333\rangle$&$[sc]_3^0[\bar{s}\bar{c}]_{\bar{3}}^1$\\
$|2612\rangle$&$f_jJ/\psi$                           &$|4343\rangle$&$[sc]_6^0[\bar{s}\bar{c}]_{\bar{6}}^1$\\
$|2622\rangle$&$[f_j]_8[J/\psi]_8$                   &$|4433\rangle$&$[sc]_3^1[\bar{s}\bar{c}]_{\bar{3}}^0$\\
$|3633\rangle$&$[sc]_3^1[\bar{s}\bar{c}]_{\bar{3}}^1$&$|4443\rangle$&$[sc]_6^0[\bar{s}\bar{c}]_{\bar{6}}^1$\\
$|3643\rangle$&$[sc]_6^1[\bar{s}\bar{c}]_{\bar{6}}^1$&$|4533\rangle$&$[sc]_3^1[\bar{s}\bar{c}]_{\bar{3}}^1$\\
$|4633\rangle$&$[sc]_3^1[\bar{s}\bar{c}]_{\bar{3}}^1$&$|4543\rangle$&$[sc]_6^1[\bar{s}\bar{c}]_{\bar{6}}^1$\\
$|4643\rangle$&$[sc]_6^1[\bar{s}\bar{c}]_{\bar{6}}^1$&&\\
\hline
\end{tabular}
\end{table}

\subsubsection{total wave function}
The total wave functions are obtained by the direct product of wave functions of orbital, spin, color and flavor wave functions.
Because we are interested in the states with quantum number $J^P=1^{-}$, there must be orbital angular momentum excitation.
The experiment suggests that the excited angular quantum number should exist in the one sub-cluster. So we follow the suggestion,
set $l_1=1, l_2=0$ or $l_1=0, l_2=1$. All the possible channels with the physical contents are listed in the table~\ref{channel}.
The subscript ``8" denotes color octet subcluster, the superscript of diquark/antidiquark is the spin of the subcluster, and the
subscript is the color representation of subcluster, $3$, $\bar{3}$, $6$ and $\bar{6}$ denote color triplet, anti-triplet,
sextet and anti-sextet.

\section{Results}
In this section, we present the numerical results of our calculation.
As a preliminary calculation, the spin-orbit interaction is not considered in the present calculation. So the states can
be classified according to the total spin $S$ of the four-quark system. All the spin of four-quark system $S=0,1,2$ can
couple with $L=1$ to give total angular momentum $J=1$.
Single channel and multi-channel coupling calculations show that no bound state can be formed.
Because of the color structures of color-octet channel in meson-meson structure and the diquark-antidiquark structure, the
system cannot fall apart directly. So in the single channel calculation, we always obtain stable energies for these channels.
To see if these states are genuine resonances or not, the real-scaling method \cite{22Tan:1981,Meng:2019fan} is employed.
In this method, the Gaussian size parameters $r_n$ for the basis functions between two sub-clusters for the color-singlet channels
are scaled by multiplying a factor $\alpha$, i.e. $r_n \rightarrow \alpha r_n$. Then, any continuum state will fall off towards
its threshold, while a compact resonant state should not be affected by the variation of $\alpha$.
\begin{figure}[htp]
\begin{center}
\centerline{\epsfxsize=9cm\epsffile{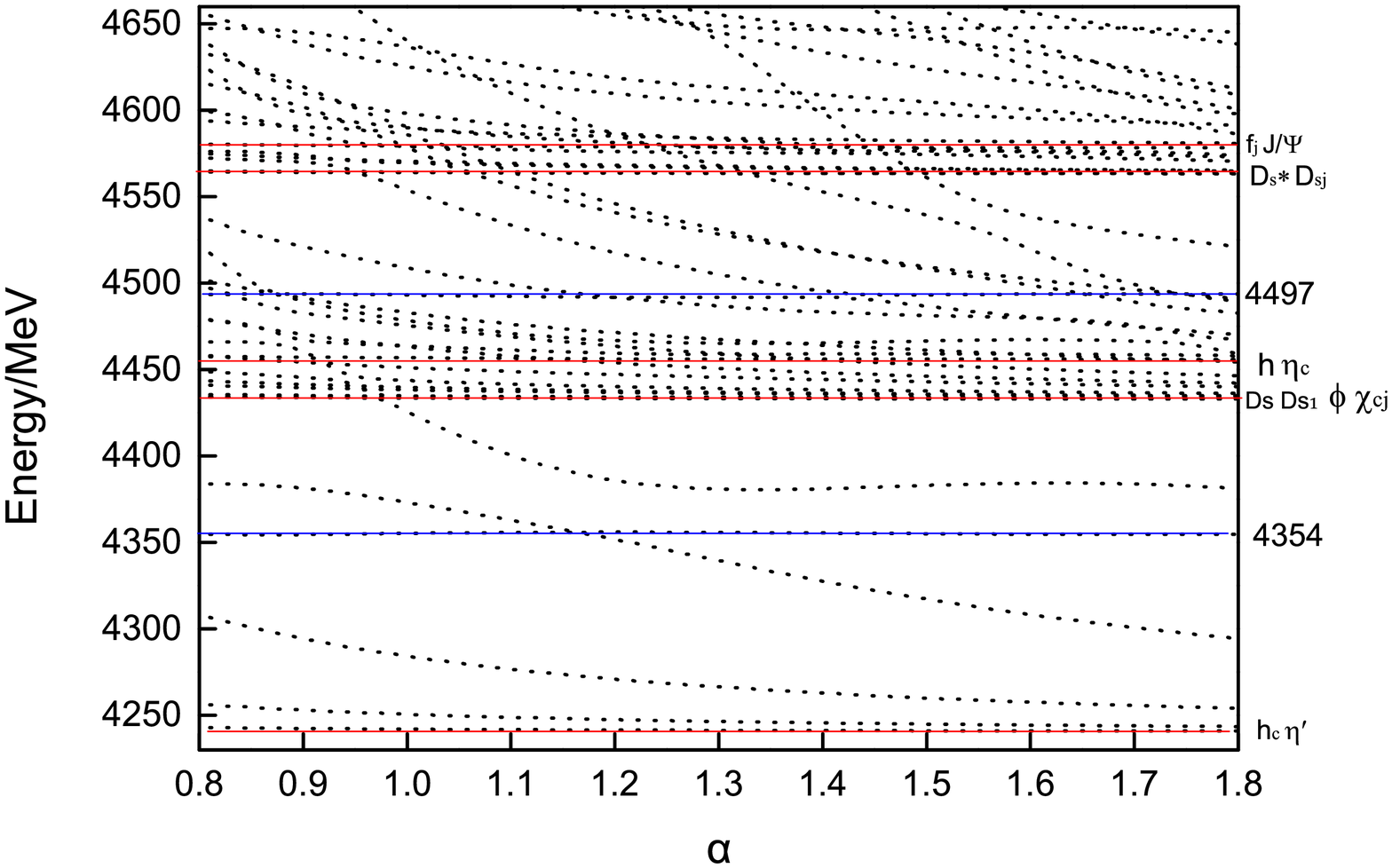}}
\caption{ Energy spectrum of $^1P_1$ states. }\label{picS0}
\end{center}
\end{figure}

\begin{figure}[htp]
\begin{center}
\centerline{\epsfxsize=9cm\epsffile{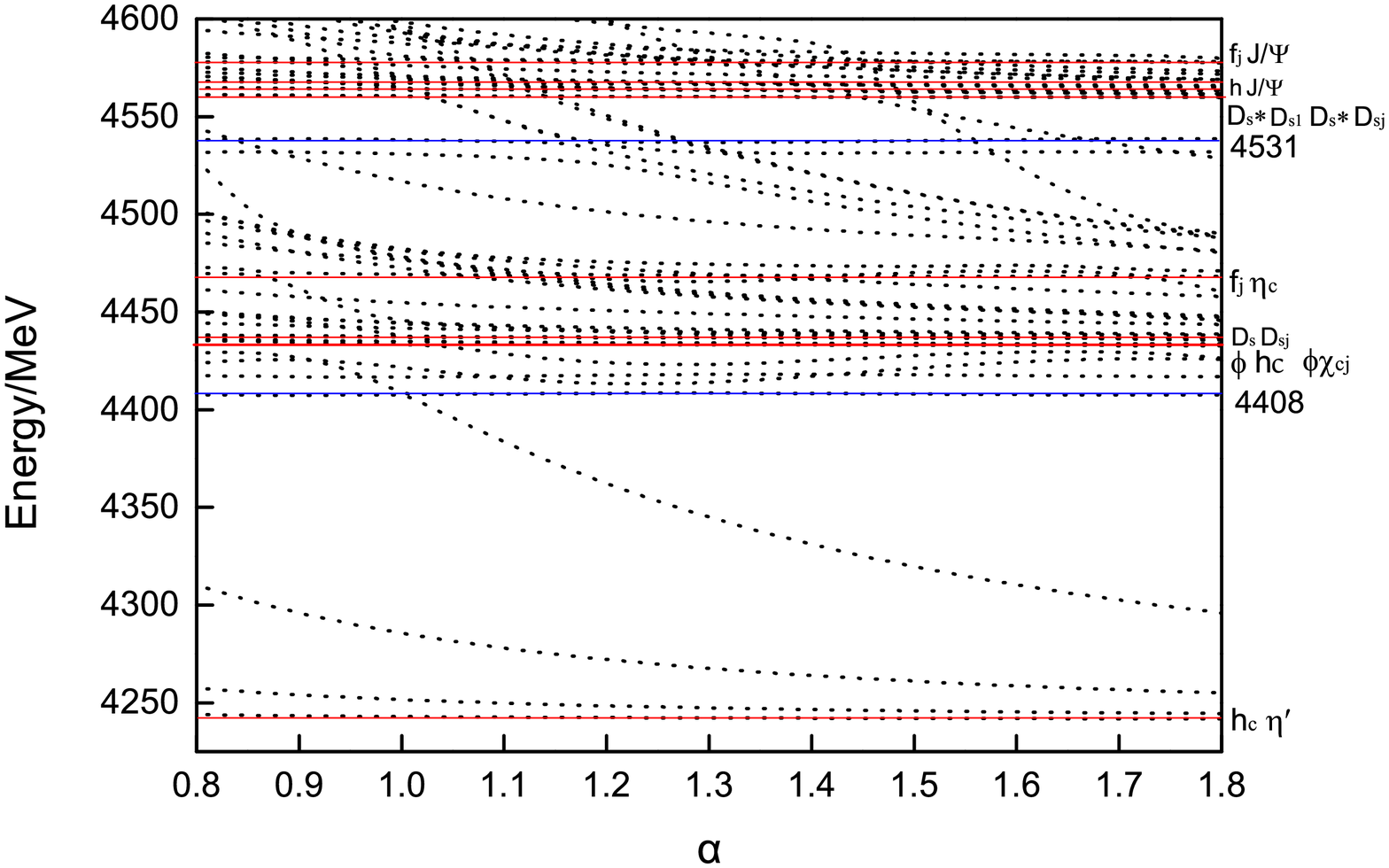}}
\caption{ Energy spectrum of $^3P_1$ states.}\label{picS1}
\end{center}
\end{figure}

\begin{figure}[htp]
\begin{center}
\centerline{\epsfxsize=9cm\epsffile{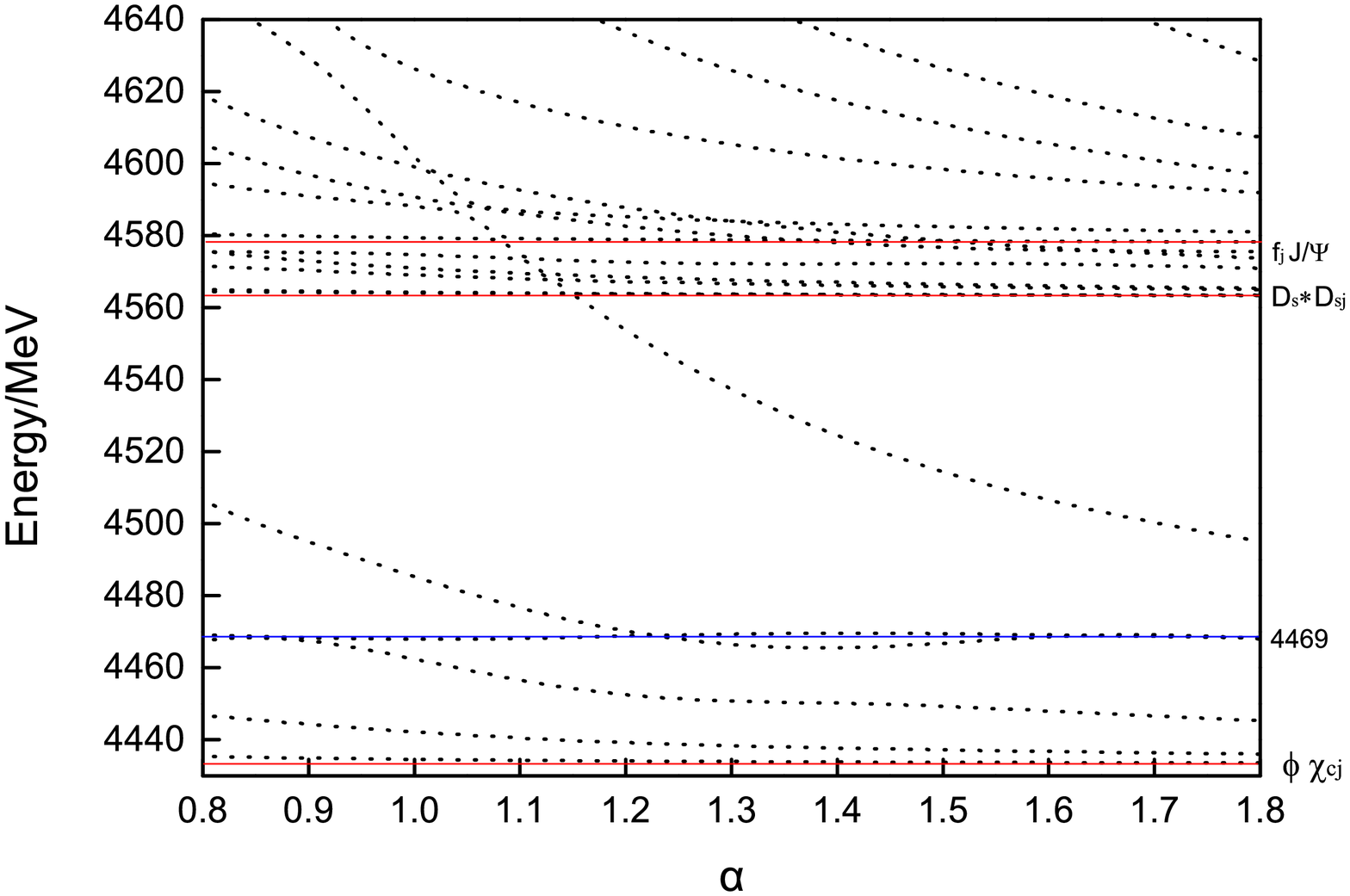}}
\caption{ Energy spectrum of $^3P_1$ states.}\label{picS2}
\end{center}
\end{figure}
The results for $S=0,1,2$ are shown in Figs.~\ref{picS0}, \ref{picS1} and \ref{picS2}. In each figure, all the thresholds
are marked with a line and physical contents. the possible resonance are also highlighted with a line and its energy,
and only the lowest one or two resonance states are displayed. From the figures, we can see that the thresholds are all
shown up with horizontal lines. Besides, there are genuine resonances, their energies are stable with the increasing $\alpha$.
For $S=0$, we obtain two resonances with energies, 4354 and 4497 MeV below 4650 MeV. For $S=1$, two resonances with energies,
4408 and 4531 MeV are shown. There is only one resonance with energy 4469 MeV below 4650 MeV.

In the quark model calculation, the masses and decay properties of hadrons can be described well. However the description cannot be
perfect. there are always some deviations. We take these deviations as our systematic errors of our calculations. From
table \ref{mesonmass}, we can see that the systematic errors in the present calculation are 60$\sim $ 100 MeV. Taking into account
of the systematic error, we find that the resonance with energy 4531 can be a candidate of the newly reported state $Y(4626)$.
\begin{table}[ht]
\centering
\caption{The average separations between any quark/antiquark pairs (unit: fm).\label{rms}}
\begin{tabular}{ccccccc}
\hline \hline
state & $r_{c\bar{s}}$  & $r_{cs}$ & $r_{c\bar{c}}$  &  $r_{s\bar{s}}$  &  $r_{\bar{s}\bar{c}}$  &  $r_{s\bar{c}}$  \\ \hline
~$R(4354)$~  & ~~0.9~~  & ~~0.9~~ & ~~0.4~~ & ~~0.9~~ & ~~0.9~~ & ~~0.9~~ \\
$R(4497)$  & 0.7  & 0.8 & 0.6 & 0.9 & 0.8 & 0.7 \\
$R(4408)$  & 0.7  & 1.4 & 1.4 & 1.4 & 1.4 & 0.7 \\
$R(4531)$  & 0.7  & 0.8 & 0.7 & 1.0 & 0.8 & 0.7 \\
$R(4469)$  & 0.9  & 0.9 & 0.4 & 0.9 & 0.9 & 0.9 \\
\hline \hline
\end{tabular}
\end{table}

To explore the structures of the resonances, the average separations between any quark/antiquark pair are calculated, the results
are shown in table \ref{rms}. From the table, we find that all the separation are not larger than 1.0 fm except for the state
$R(4408)$, so these states are compact objects. For the state $R(4408)$, we have small separations $r_{c\bar{s}}$ and $r_{\bar{s}c}$, 
and a little large separations $r_{cs}$, $r_{c\bar{c}}$, $r_{s\bar{s}}$ and $r_{\bar{s}\bar{c}}$, so it is a molecule. 
For the state $R(4531)$, a candidate of the state $Y(4626)$, all the separation are around 0.8 fm, so it is a compact tetraquark state.
The wavefunction of the state supports the picture, where the configurations with colorful subclusters dominant. 

\section{Summary}

In the framework of the chiral constituent quark model, we study systematically $J^P=1^{-}$ $c\bar{s}s\bar{c}$ states. 
Two different structures, meson-meson structure and diquark-antidiquark, with all possible color, flavor, spin configurations 
are taken into account. In the absence of spin-orbit interaction, we found that there is no bound state for this system.
However, the resonances are possible. To distinguish the genuine resonances from the discretized scattering states, the
real-scaling method is employed. The calculations show that there are five resonance states with $J^P=1^-$ and $S=0,1,2$.
One state with molecular structure, and other states are all have a compact structure. The newly observed state $Y(4620)$ 
can be described as a $c\bar{s}c\bar{c}$ compact tetraquark state. 

When the spin-orbit and tensor interactions are included, all the states with $S=0$, $S=1$ and $S=2$ will be mixed up. 
Clearly, further calculation is expected. Whether the state can survive after invoking spin-orbit and tensor interactions? 
If these states survive, the decay widths have to be calculated to check the compatibility with the experimental data. 
Are there other explanations of the state $Y(4620)$? These are our future work.

\acknowledgments{This work is supported partly by the National Natural Science Foundation of China under
Contract Nos. 11675080, 11175088 and 11535005.}

\end{document}